\documentstyle[12pt]{article}
\def\prl{Phys.\ Rev.\ Lett.\ }

\def\prb{Phys.\ Rev.\ B }
\begin{document}
\setlength{\baselineskip}{.375in}
\newcommand{\wn}{\omega_n}
\begin{flushright}
NSF-ITP-92-54
\end{flushright}
\vspace{0.7in}
\begin{center}
\Large\bf
Universal quantum-critical dynamics of two-dimensional antiferromagnets\\
\vspace{0.75in}
\normalsize\rm
Subir Sachdev and Jinwu Ye\\
\vspace{0.2in}
\em
Institute for Theoretical Phyics, University of California,
Santa Barbara CA 93106\\
and\\
Center for Theoretical Physics, P.O. Box 6666,
Yale University, New Haven, CT 06511
\end{center}
\vspace{0.7in}
\rm
The universal dynamic and static properties of two dimensional
antiferromagnets in the vicinity of a zero-temperature
phase transition from long-range magnetic order to a quantum
disordered phase are studied.
Random antiferromagnets with both N\'{e}el and spin-glass
long-range magnetic order are considered. Explicit quantum-critical
dynamic
scaling functions are computed in a $1/N$ expansion to
two-loops for certain non-random, frustrated
square lattice antiferromagnets.
Implications for neutron scattering experiments
on the doped cuprates are noted.
\begin{verbatim}
July 1992
PACS Nos 75.10.J, 75.50.E, 05.30
\end{verbatim}

\newpage
Recently, there have been a number of fascinating and
detailed experiments~\cite{birg,gabe,kagome}
on layered antiferromagnets (AFMs)
close to a zero-temperature ($T$) phase transition at which
magnetic long-range-order (LRO) vanishes.
The most prominent among these are
the cuprates~\cite{birg,gabe}, which, upon doping with a
small concentration of holes, lose their long-range N\'{e}el
order and undergo a transition to a
$T=0$ phase with magnetic, long-range spin-glass order; at a
larger doping there is presumably a second transition to a
quantum-disordered (QD) ground state.
There have also been
low-$T$ experiments on layered AFMs on frustrated
lattices~\cite{kagome}, which have at most a small
ordering moment.
A remarkable feature of the measured dynamic susceptibilities of
these AFMs is
that overall frequency scale of the spin excitation spectrum is given
simply by
the absolute temperature. In particular, it appears to be
independent of all microscopic energy scales
{\em e.g.\/} an antiferromagnetic exchange constant.

In this paper, we show that this anomalous dynamics
is a very general property of finite $T$,
`quantum-critical' (QC)~\cite{sudip}
spin fluctuations near the initial onset of a
$T=0$ QD phase.
We
present the first calculation of
universal, QC
dynamic scaling
functions in 2+1 dimensions; these will be calculated for
a model system - non-random, frustrated
two-dimensional Heisenberg AFMs with a vector
order-parameter.
Quenched randomness will be shown to
be a relevant perturbation to the clean system, and must be included in
any comparison with experiments.
Scaling forms for the dynamic susceptibility in random
AFMs will be presented, and
exponent (in-)equalities will be discussed.

QC dynamic scaling functions can also be studied in
other dimensions. In 1+1 dimensions the exact scaling
functions can be obtained by a simple argument based on
conformal invariance~\cite{cardy}; most 3+1
dimensional models are in the upper-critical dimension and
we expect the scaling functions to be free-field with
logarithmic
corrections. This leaves 2+1 dimensions, which is studied here
for the first time
in the context of AFMs:
however, our results are more general, and
should also be applicable to other
phenomena like the
superconductor-insulator transition~\cite{matthew}.

Most of our discussion will be in the context of
the following quantum AFMs:
\begin{equation}
{\cal H} = \sum_{i,j} J_{ij} {\bf S}_{i} \cdot {\bf S}_{j}
\end{equation}
where ${\bf S}_i$ are quantum spin-operators on the sites, $i$, of a
two-dimensional lattice, and the $J_{ij}$ are a set of possibly
random, short-range antiferromagnetic exchange interactions.
The lightly-doped cuprates are insulating at $T=0$, suggesting a
model with completely localized holes: a specific form of ${\cal H}$
with frustrating interactions was used by Gooding and
Mailhot~\cite{gooding} and yielded reasonable results on the doping
dependence of the $T=0$ correlation length. Models with mobile holes
have also been considered~\cite{longpap} and the results will
be noted later.

Two different classes of ground states of ${\cal H}$ can be distinguished:
({\em i\/}) states with magnetic LRO $\langle {\bf S}_i
\rangle = {\bf m}_i$ and ({\em ii\/}) QD states which
preserve spin-rotation invariance $\langle {\bf S}_i \rangle = 0$.
Further, we will distinguish between two
different types of magnetic LRO: ($A$)
N\'{e}el LRO in which case ${\bf m}_i \sim e^{i \vec{Q} \cdot
\vec{R}_i}$ with $\vec{Q}$ the N\'{e}el ordering wavevector
and ($B$) spin-glass LRO in which case ${\bf m}_i$ can have an
arbitrary dependence on $i$, specific to the particular realization of
the randomness. The lower critical dimension of the Heisenberg
spin-glass~\cite{young_binder}
may be larger than 3 - in this case
the spin-glass LRO will not survive to any finite $T$, even in
the presence of a coupling between the layers. This, however, does
not preclude the existence of spin-glass LRO at $T=0$.

Consider now a $T=0$ phase transition between the magnetic LRO and
the QD phases, induced by varying a coupling constant
$g$ (dependent on the ratio's of the $J_{ij}$ in ${\cal H}$)
through a critical value $g=g_c$, where there is
a diverging correlation length $\xi \sim |g - g_c
|^{-\nu}$. For N\'{e}el
LRO, $1/\xi$ is the width of the peak in the spin structure-factor at
the ordering wavevector $\vec{Q}$. For spin-glass LRO, there
is no narrowing of the structure-factor, and $\xi$ is instead a
correlation length associated with certain four-spin correlation
functions~\cite{young_binder}. At finite $T$ we can
define a thermal length $\xi_T \sim T^{-1/z}$ ($z$ is the
dynamic critical exponent) which is the scale at which deviations
from $T=0$ behavior are first felt. The QC region is
defined by the inequality $\xi_T < \xi$ (Fig. 1); in this case the spin-system
notices the finite value of $T$ {\em before\/} becoming sensitive to
the deviation of $g$ from $g_c$, and the dynamic spin
correlations will be found to be remarkably universal.

We consider first case $(A)$ - a phase transition from N\'{e}el LRO to
a QD phase; such transitions can occur for both random and non-random
${\cal H}$. At $T=0$ the static spin susceptibility, $\chi$,
will have a divergence at $g=g_c$ and wavevector $\vec{q} = \vec{Q}$:
$\chi(\vec{q} = \vec{Q}, \omega = 0 ) \sim |g-g_c|^{-\gamma}$ with
$\gamma = (2 - \eta) \nu$. The form of the $\vec{q}, \omega, T$
dependent susceptibility in the QC region
can be obtained simply by finite-size scaling: $\xi_T$ acts a
finite-size in the imaginary-time direction for the quantum system at its
critical point~\cite{sudip,matthew} and hence implies the scaling form
\begin{equation}
\chi ( \vec{q} , \omega ) = \frac{a_1}{T^{(2-\eta)/z}} \Phi \left(
\frac{a_2 |\vec{q}-\vec{Q}|}{ T^{1/z}}, \frac{\hbar \omega}{k_B T} \right)
\label{scale_q}
\end{equation}
where $a_1 , a_2$ are non-universal constants, and $\Phi$ is a
universal, complex function of both arguments. The deviations from
quantum-criticality lead to an additional dependence of $\Phi$ on
$\xi_T / \xi$: this number is small in the QC
region and has been set to 0.
Also of experimental interest is the
local dynamic susceptibility $\chi_L ( \omega ) = \int d\vec{q}~
\chi ( \vec{q}, \omega ) \equiv \chi_L^{\prime} + i
\chi_L^{\prime\prime}$, with real (imaginary) part $\chi_L^{\prime}$
($\chi_L^{\prime\prime}$) (Note $d \vec{q} \equiv d^2 q / (4
\pi^2 )$).
As $\chi^{\prime} \sim |\vec{q}
- \vec{Q}|^{-2+\eta}$ for $|\vec{q}-\vec{Q}| \gg \omega^{1/z}, T^{1/z}$,
the real-part of the $\vec{q}$ integral
is dominated by its singular piece only if $\eta <
0$. However $\chi_L^{\prime\prime}$ will involve only on-shell
excitations, and the imaginary part of the $\vec{q}$ integral is
expected to be convergent in the ultraviolet for both signs of
$\eta$. Thus the leading part of $\chi_L^{\prime\prime}$ will always
obey the scaling form
\begin{equation}
\chi_L^{\prime\prime} (\omega ) = a_3~ |\omega|^{\mu}~ F \left(
\frac{\hbar \omega}{k_B T} \right)
\label{scale_loc}
\end{equation}
\begin{equation}
\mbox{with}~~~~~~\mu = \eta/z,
\end{equation}
$F (y) = y^{-\mu} \int d
\vec{x}~ \mbox{Im} \Phi ( \vec{x}, y )$ a universal function, and
$a_3$ a non-universal number.
$\chi_L^{\prime}$ also has a part obeying an
identical scaling form which is dominant
only if $\eta < 0$.
As we expect $\chi_L^{\prime\prime} \sim \omega$ for small $\omega$,
we have the limiting forms $F \sim \mbox{sgn}(y)
|y|^{1-\mu}$ for $y \ll 1$ and $F
\sim \mbox{sgn}(y)$ for $y \gg 1$. Note that all the non-universal
energy scales only appear in the prefactor $a_3$ and the frequency
scale in $F$ is determined solely by $T$.

Now the other case $(B)$: the transition from spin-glass LRO to a QD phase. We
do not expect singular behavior as a function of $\vec{q}$ because
the spin-condensate ${\bf m}_i$ is a random function of $i$; therefore the
scaling form (\ref{scale_q}) will {\em not} be obeyed. However
the local susceptibility $\chi_L ( \omega_n ) \equiv \int_0^{1/(k_B T)} d
\tau e^{i \omega_n \tau} C(\tau)$, $C(\tau) =
\overline{ \langle {\bf S}_i (0) \cdot {\bf
S}_i ( \tau ) \rangle }$ (where $\omega_n$ ($\tau$) is a Matsubara
frequency (time) and the bar represents average over sites $i$)
will be quite sensitive to spin-glass LRO. In the spin-glass phase at $T=0$
$\lim_{\tau\rightarrow\infty} C(\tau ) =
\overline{ {\bf m}_i^2 } > 0$~\cite{young_binder}. In the QD phase,
numerical studies of random, spin-1/2
AFMs~\cite{bhatt_lee} suggest that at $T=0$, $C(\tau) \sim 1/\tau^{1-\alpha}$,
$\alpha > 0$, for large $\tau$. At the critical point $g=g_c$ and
$T=0$ we therefore expect the intermediate scaling behavior with
$C(\tau) \sim 1/\tau^{1+\mu}$,
$\chi_L^{\prime} \sim
\chi_L^{\prime\prime} \sim |\omega|^{\mu}$ and $-1 < \mu
< 0$.
In the QC region the scaling form (\ref{scale_loc}) for
$\chi_L$ continues to be valid, despite the inapplicability of
(\ref{scale_q}). The limiting forms for $F$ are as in $(A)$, although
the value $\mu$ is different:
the Edwards-Anderson order-parameter obeys $\overline{ {\bf
m}_i^2 } \sim |g - g_c|^{\beta}$ for $g < g_c$ - connecting the
form of $\chi_L$ in the spin-glass phase to the critical point, we get
\begin{equation}
\mu = - 1 + \beta/(z\nu)
\end{equation}

We now consider various model systems for which exponents and/or
scaling functions have be computed.

We consider first a transition from N\'{e}el LRO to a QD phase in a
non-random spin-1/2 square lattice AFM with {\em e.g.\/} first ($J_1$)
and
second ($J_2$) neighbor interactions~\cite{gsh,spn1}.
As has been discussed in great detail elsewhere~\cite{spn1},
spin-Peierls order appears in the QD phase in this case
(and in all other
non-random AFMs with commensurate, collinear, N\'{e}el LRO).
We now argue that the
two-spin, QC dynamic scaling
functions are not sensitive to the spin-Peierls fluctuations, and
one may use an effective action for only the N\'{e}el order.
It was found in the large $M$ calculations for $SU(M)$ AFMs that the
asymptotic decay of the spin-Peierls correlations are governed by a scale
$\xi_{SP}$, which is much larger (for $M$ large) than the scale,
$\xi$, governing the decay of the N\'{e}el order~\cite{spn1,murthy}.
The two scales are related by $\xi_{SP} = \xi^{4M\rho_1}$,
where $\rho_1$ is a critical exponent given by $\rho_1 = 0.062296$ to
leading order in $1/M$~\cite{murthy}. D.S. Fisher~\cite{daniel}
has noted that this is reminiscent of three-dimensional statistical
models with a `dangerously irrelevant' perturbation: {\em e.g.\/}
the $d=3$
classical $XY$ model with a cubic
anisotropy~\cite{cubic} has a phase-transition in the pure XY-class,
but the `irrelevant' cubic anisotropy becomes important in the
low-$T$
phase at distances larger than $\xi_{XY}^{\psi}$ ($\psi > 1$).
By analogy, we may conclude that the spin-Peierls
fluctuations are irrelevant at the critical fixed point governing the
quantum phase transition, and relevant only at the strong-couping
fixed point which governs the nature of the QD phase.

It has been argued in Ref.~\cite{sudip} that the dynamics of the
N\'{e}el order-parameter is well described by an $O(3)$ non-linear
sigma ($NL\sigma$) model in the renormalized classical region
(Fig.~1).
The gist of the above arguments is that this mapping continues to be
valid in the QC region - but not any further into the QD
phase! We have computed properties of the QC region by
a $1/N$ expansion on a $O(N)$ $NL\sigma$ model:
\begin{equation}
S_{\hat{n}} = \frac{1}{2g} \int d\tau d^2 r \left[ (\nabla \hat{n})^2 +
\frac{1}{c^2} \left( \frac{\partial \hat{n}}{\partial \tau} \right)^2
\right] ~~~~~\hat{n}^2 = 1
\label{nls}
\end{equation}
were $\hat{n}$ is a real $N$-component spin field, and $c$ is a
spin-wave velocity.
The saddle-point equations of the large $N$ expansion~\cite{polyakov}
were solved and the correlation functions
were shown to satisfy the scaling forms
(\ref{scale_q},\ref{scale_loc})
to order $1/N$ (two-loops).
We determined the values of $\Phi(x,y)$ for real frequencies $y$
by analytically continuing the Feynman graphs and subsequently
numerically
evaluating
the integrals.
The numerical computations required the equivalent of
40 hours of vectorized supercomputer time.

Our results for $\mbox{Im} \Phi$ and $F$ for $N=3$ are
summarized in Figs.~2,3. The transition has the exponent $z=1$
which fixes the constant $a_2 = \hbar c/k_B$ in
Eqn.~(\ref{scale_q}).
We normalized $\Phi ( x, y) $ such that $\partial \Phi^{-1}/\partial x^2
|_{0,0} = 1$. Analytic forms for $\Phi$ can be obtained in various
regimes.
We have
\begin{equation}
\mbox{Re} \Phi^{-1} = C_Q^{-2} + x^2 + \ldots~~~~\mbox{$x,y$ small}
\end{equation}
The universal number $C_Q$, to order $1/N$, is:
\begin{equation}
C_Q^{-1} = \Theta \left( 1 + 0.22/N \right) ~~~;~~~
\Theta = 2 \log\left((1 + \sqrt{5})/2 \right)
\end{equation}
$\mbox{Im} \Phi$ has a singular behavior for $x,y$ small:
$\mbox{Im} \Phi (x=0, y) \sim \exp(-3 \Theta^2 /(2 |y|))/N$
while $\mbox{Im} \Phi ( y < x) \sim y \exp(-3 \Theta^2 /(2|x|))
/N$.
With either $x,y$ large, $\Phi$ has the form
\begin{equation}
\Phi = D_{Q}\/ (x^2 - y^2)^{-1+\eta/2} + \ldots~~~~~;~~~~~
D_Q = 1 - 0.3426/N.
\end{equation}
The exponents $\mu,\eta$ have
the known~\cite{abe} expansion $\mu=\eta = 8/(3 \pi^2
N) - 512/(27\pi^4 N^2 )> 0 $.
The scaling function for the local susceptibility,
$F(y)$, has the limiting forms
\begin{equation}
F(y) = \mbox{sgn}(y)\frac{0.06}{N} |y|^{1-\eta} ~~~
\mbox{$y \ll 1$}~~~;~~~
F(y) = \mbox{sgn}(y) \frac{D_Q}{4} \frac{\sin ( \pi \eta /2 )}{\pi \eta /2}
{}~~~\mbox{$y \gg 1$}
\end{equation}
As $\eta$ is small, $F$ is almost linear at small $y$.
At $N=\infty$, $F = \mbox{sgn} (y) \theta( |y|-\Theta )/4$.

We now study ${\cal H}$ with quenched randomness. The simplest
model adds a small fluctuation in the $J_1$ bonds of the $J_1 - J_2$
model above: {\em i.e.\/} $J_1 \rightarrow J_1 + \delta
J_1$ where $\delta J_1$ is random, with r.m.s. variance $\ll
J_1$, ensuring that a N\'{e}el LRO to QD transition will continue to
occur. However the transition will not be described by the `pure'
fixed point as $\nu_{\mbox{pure}} = 0.705 \pm 0.005$~\cite{nupure}
and thus violates the
bound $\nu > 2/d = 1$ required of phase
transitions in random systems~\cite{chayes}.  At long
wavelengths we expect the spin fluctuations
to be described by the $NL\sigma$ model, $S_{\hat{n}}$,
(\ref{nls}) with random, space-dependent, but time-independent,
couplings $g,c$. A soft-spin version of $S_{\hat{n}}$ with random
couplings in $d=4-\epsilon-\epsilon_{\tau}$ space dimensions and
$\epsilon_{\tau}$ time dimensions has been examined in a double
expansion in $\epsilon, \epsilon_{\tau}$~\cite{dbc}. The expansion is
poorly-behaved, and for the case of interest here ($N=3$, $\epsilon=1$,
$\epsilon_{\tau} = 1$) the random fixed-point has the exponent
estimates  $\eta = -0.17$,
$z=1.21$, $\nu = 0.64$, $\mu=-0.15$.
Note ({\em i\/}) $\mu,\eta$ are negative, unlike
the pure fixed point and ({\em ii\/}) $\nu$ is smaller than $2/d$
suggesting large higher-order corrections.

Consider next a ${\cal H}$ on the square lattice with only $J_1$
couplings, but with a small concentration of static, spinless holes
on the vertices; this model will
display a N\'{e}el LRO to QD transition at a critical
concentration of holes. In the coherent-state path-integral
formulation of the pure model, each spin contributes a Berry phase
which is almost completely canceled in the continuum limit between
the contributions of the two sublattices~\cite{berry}. The model
with holes will have large regions with unequal numbers of spins on
the two sublattices: such regions will contribute a Berry phase which
will almost certainly be relevant at long distances. Therefore the
field theory of Ref.~\cite{dbc} is not expected to describe the
N\'{e}el LRO to QD transition in this case. A cluster expansion in
the concentration of {\em spins} has recently been carried out by Wan
{\em et. al.}~\cite{wan}: and yields the exponents $\eta = -0.6$, $z=1.7$,
$\nu = 0.8$, $\mu=-0.35$. Note again that $\mu,\eta < 0$, although the
violation
of $\nu > 2/d$ suggests problems with the series extrapolations.

Finally, we have also
considered~\cite{longpap} the consequences of mobile holes
in a non-random AFM. The
spin-waves and holes were described by the
Shraiman-Siggia~\cite{boris}  field theory.
Integrating out the fermionic holes, led to a spin-wave self-energy
$\Sigma_{\hat{n}} \sim a_1 |\vec{q}-\vec{Q}|^2 + a_2 \omega_n^2 + \ldots$,
($a_1 , a_2$ constants) at $g=g_c$, $T=0$; non-analytic $|\omega_n|$ terms
appear
only with higher powers of $|\vec{q}-\vec{Q}|, \omega_n$ indicating
that the
N\'{e}el LRO to QD transition has the same leading
critical behavior as that in the undoped, non-random $J_1-J_2$ model
above. The exponents and scaling functions are identical, but the
corrections to scaling are different.

To conclude, we discuss
implications for neutron scattering experiments in the
doped cuprates~\cite{birg,gabe}.
The significant low $T$ region with a $T$-independent
width of the spin structure-factor indicates that the experiments can
only be in the QC region (Fig.~1) of a $T=0$ transition
from spin-glass LRO to QD:
the diverging spin-glass correlation length
will then not be apparent
in the two-spin correlations. The numerical results of
Ref.~\cite{gooding} also indicate that, in the absence of a coupling
between the planes, a spin-glass phase will appear at any non-zero
doping. The experimental $\chi_L^{\prime\prime}$ has been fit to a
form $I(|\omega|) F(\hbar\omega / k_B T)$~\cite{birg} which is
compatible with the theoretical QC result
(\ref{scale_loc}) if $I \sim |\omega|^{\mu}$. A fit to this form for
$I$ in $La_{1.96} Sr_{0.04} Cu O_4$ yielded $\mu = -0.41 \pm 0.05$
with all the predited points within the experimental error-bars~\cite{bernie}.
As it appears that only random models have $\mu < 0$,
it is clear that the effects of randomness are experimentally crucial,
confirming the theoretical prediction of their relevance.
Further theoretical work on the QC dynamics of random
quantum spin models is clearly called for.

We thank B.~Keimer, G.~Aeppli, R.N.~Bhatt, D.S., M.E. and
M.P.A.~Fisher, B.I.~Halperin,
D.~Huse, N.~Read, and A.P.~Young for useful discussions.
This research
was supported
by NSF Grants No. DMR 8857228 and PHY89-04035,
and the A.P. Sloan Foundation.

\newpage
\section*{\large\bf Figure Captions}
\begin{enumerate}
\item
Phase diagram of ${\cal H}$ (after Ref.~\cite{sudip}). The magnetic LRO
can be either spin-glass or N\'{e}el, and is present only at $T=0$.
The boundaries of the QC region are $T \sim |g-g_c|^{z\nu}$.
For non-random ${\cal H}$ which have commensurate, collinear, N\'{e}el
LRO for $g<g_c$,
all of the QD region ($g>g_c$) has spin-Peierls order at $T=0$-this order
extends to part of the QD region at finite $T$.
\item
The imaginary part of the universal susceptibility in the QC region,
$\Phi$, as a function of $x = \hbar c q/(k_B T)$ and $y = \hbar \omega
/ (k_B T) $ for a non-random square lattice AFM which undergoes a $T=0$
transition from N\'{e}el LRO to a QD phase. The results have been
computed in a $1/N$ expansion to order $1/N$ and evaluated for $N=3$.
The two-loop diagrams were analytically continued to real frequencies
and
the integrals then evaluated numerically.
The shoulder on the peaks is due to a threshold towards three
spin-wave decay.
\item
The imaginary part of the universal local susceptibility, $F$, for the
same
model as in Fig~2. We have $F(y) = y^{-\mu} \int d\vec{x} ~\mbox{Im} \Phi (
\vec{x},
y )$.
The oscillations at large $y$ are due to a finite step-size in the
momentum integrations.
\end{enumerate}
\end{document}